\RequirePackage{filecontents}

\RequirePackage{snapshot}
\documentclass[%
secnumarabic,
twoside,
floatfix,
a4paper,
superscriptaddress,
groupedaddress,
amsmath,amssymb,amsfonts,amsbsy,
aps,reprint,twocolumn,notitlepage,longbibliography]{preprint}

\makeatletter
\def\lrbox#1{%
  \edef\reserved@a{%
    \endgroup
    \global\setbox#1\hbox{%
      \begingroup\aftergroup}%
    \def\noexpand\@currenvir{\@currenvir}%
    \def\noexpand\@currenvline{\on@line}}%
  \reserved@a
  \@endpefalse
  \color@setgroup
  \ignorespaces}
\def\endlrbox{\unskip\color@endgroup}
\makeatother

\renewcommand\documentclass[2][]{}
\newcommand{\changefont}[3]{}
\newcommand{\journal}[1]{}
\newbox\mybox
\newenvironment{frontmatter}{}{\maketitle}
\let\ead=\email

\AtBeginDocument{%
  \IfFileExists{date.tex}{%
    \input{date.tex}

  }
}

\newcommand{\corref}[2][]{}
\newcommand{\cortext}[2][]{}
\makeatletter
\AtBeginDocument{
  \def\@linkcolor{blue}
  \def\@anchorcolor{blue}
  \def\@citecolor{blue}
  \def\@filecolor{blue}
  \def\@urlcolor{blue}
  \def\@menucolor{blue}
  \def\@pagecolor{blue}
  \def\bibsection{%
   \par
   \begingroup
    \baselineskip26\p@
    \bib@device{\hsize}{72\p@}%
   \endgroup
   \nobreak\@nobreaktrue
   \addvspace{19\p@}%
  }%
}
\makeatother
\pdfoutput=1
\documentclass[final,5p,times,twocolumn,sort&compress]{elsarticle}

\usepackage{graphicx}

\graphicspath{{figs/}}

\usepackage[utf8]{inputenc}
\usepackage[T1]{fontenc}
\usepackage{txfonts}
\usepackage{tgtermes}
\usepackage[altg]{qtxmath}
\usepackage{bm}
\usepackage{braket}
\usepackage{microtype}
\usepackage{amsmath}
\usepackage{amssymb}
\usepackage{amsfonts}
\usepackage[%
pdfborder={0 0 0}, 
breaklinks=true,
colorlinks=true,
pdfauthor={Anton W{\"o}llert, Heiko Bauke, Christoph H. Keitel},
pdftitle={Multi-pair states in electron-positron pair creation}]
{hyperref}

\newcommand{\norm}[1]{\vert #1 \vert}
\newcommand{\op}[1]{\hat{#1}}
\renewcommand*{\vec}[1]{\bm{#1}}
\newcommand*{\D}{\mathrm{d}}

\newcommand*{\I}{\mathrm{i}}
\newcommand*{\e}{\mathrm{e}}
\renewcommand*{\Re}[1]{\operatorname{Re}{#1}}

\newcommand{\tin}{t_{\mathrm{in}}}
\newcommand{\tout}{t_{\mathrm{out}}}

\newcommand*{\trans}[1]{#1^{\mathsf{T}}}
\newcommand*{\herm}[1]{#1^{\dagger}}
\newcommand*{\HD}{\mathcal{H}}
\newcommand*{\phiin}[1][\pm]{{}_{#1}\varphi}
\newcommand*{\phiout}[1][\pm]{{}^{#1}\varphi}
\newcommand*{\Ein}[1][\pm]{{}_{#1}\varepsilon}
\newcommand*{\Eout}[1][\pm]{{}^{#1}\varepsilon}
\newcommand*{\Gud}[2]{G({}^#1|{}_#2)}
\newcommand*{\GudInv}[2]{G^{-1}({}^#1|{}_#2)}

\newcommand*{\wmn}{\omega(\overset{+}{m} \overset{-}{n}|0)}
\newcommand*{\wmnin}{\smash[t]{\wmn}}
\newcommand*{\wmni}{\omega(\overset{+}{m_i} \overset{-}{n_i}|0)}

\newcommand*{\mnket}{\ket{N_{\{m,n\}}}}
\newcommand*{\mnbra}{\bra{N_{\{m,n\}}}}
\newcommand*{\cmn}{{c_{\{m,n\}}}}
\newcommand*{\cpmn}{{{c'}_{\kern-0.4ex\{m,n\}}}}
\newcommand*{\outbra}{\bra{\text{out}}}
\newcommand*{\outket}{\ket{\text{out}}}

\begin{document}

\begin{frontmatter}
\title{Multi-pair states in electron--positron pair creation}

\author{Anton W{\"o}llert}%
\ead{woellert@mpi-hd.mpg.de}

\author{Heiko Bauke\corref{corr}}
\ead{heiko.bauke@mpi-hd.mpg.de}

\author{Christoph H.~Keitel\corref{}}
\address{Max-Planck-Institut f{\"u}r Kernphysik, Saupfercheckweg 1, 69117 Heidelberg, Germany}

\cortext[corr]{Corresponding author}

\date{\today}

\begin{abstract}
  Ultra strong electromagnetic fields can lead to spontaneous creation
  of single or multiple electron--positron pairs.  A quantum field
  theoretical treatment of the pair creation process combined with
  numerical methods provides a description of the fermionic quantum field
  state, from which all observables of the multiple electron--positron
  pairs can be inferred.  This allows to study the complex
  multi-particle dynamics of electron--positron pair creation in-depth,
  including multi-pair statistics as well as momentum distributions and
  spin.  To illustrate the potential benefit of this approach, it is
  applied to the intermediate regime of pair creation between
  nonperturbative Schwinger pair creation and perturbative multiphoton
  pair creation where the creation of multi-pair states becomes
  nonnegligible but cascades do not yet set in. Furthermore, it is
  demonstrated how spin and helicity of the created electrons and
  positrons are affected by the polarization of the counterpropagating
  laser fields, which induce the creation of electron--positron pairs.
\end{abstract}
\end{frontmatter}


\allowdisplaybreaks
\enlargethispage{2.5ex}

\section{Introduction}
\label{sec:intro}

Relativistic quantum field theory predicts the possibility of
electron--positron pair creation from the vacuum in the presence of a
strong electromagnetic field.  Initiated by the pioneering works by
Sauter, Heisenberg, and Euler
\cite{sauter1931verhalten,heisenberg1936folgerungen} many theoretical
\cite{walter1985effective,kluger1992fermion,schmidt1998quantumkinetic,ruf2009pair,cheng2009creation,bulanov2010multiple,dipiazza2012extremely,kohlfurst2014effective,pike2014photon,kasper2014fermion,wollert2015spin,Meuren:Hatsagortsyan:etal:2015:High-Energy_Recollision}
and experimental
\cite{ahmad1997search,burke1997positron,bamber1999studies} efforts have
been undertaken to study pair creation; see
Refs.~\cite{ehlotzky2009fundamental,dipiazza2012extremely} for recent
reviews.  Spontaneous pair creation by static fields is expected to set
in at the Schwinger critical field strength of
$E_\mathrm{S}=1.3\times10^{18}\,\text{V/m}$, which cannot be reached
even by the strongest laser facilities available today.  However, pair
creation may be assisted by time- and space-dependent electromagnetic
fields.  Novel light sources envisage to provide field intensities in
excess of $10^{20}\,\mathrm{W/cm^2}$ and field frequencies in the
\mbox{X-ray} domain
\cite{altarelli2006xfel,yanovsky2008ultra,mcneil2010xray,emma2010first,mourou2012exawatt}.
The ELI-Ultra High Field Facility, for example, aims to reach
intensities exceeding even $10^{23}\,\mathrm{W/cm^2}$ corresponding to a
field strength of about $10^{15}\,\mathrm{V/m}$, which may be sufficient
to observe pair creation
\cite{ringwald2001pair,alkofer2001pair,Narozhny:Bulanov:etal:2004:On_e,blaschke2006pair,bell2008possibility,blaschke2009dynamical,dunne2009newstrong}.

Pair creation in time-dependent electromagnetic fields can be
characterized via the classical nonlinearity parameter
$\xi=eE/(m_0c\omega)$ as nonperturbative Schwinger pair creation
($\xi\gg1$) or perturbative multiphoton pair creation ($\xi\ll1$), where
$m_0$ denotes the electron mass, $c$ the speed of light, $e$ the
elementary charge, $E$ the electric field's peak strength, and $\omega$
its angular frequency.  Both regimes are accessible by different
analytical methods.  Experimentally the nonperturbative Schwinger regime
may be realized by high-intensity optical lasers and has attracted a
considerable amount of theoretical research, e.\,g., predicting
pair-creation cascades by semi-classical methods
\cite{fedetov2010limitations,Bulanov:Esirkepov:etal:2010:Schwinger_Limit,elkina2011qed,Ridgers:Brady:etal:2012:Dense_Electron-Positron,king13photon,mironov2014collapse,bashmakov2014effect,gonoskov2015extended}.
In this regime, pair-creation can be understood as a tunneling process
\cite{brezin1970pair,wollert2015relativistic}, which is exponentially
suppressed for subcritical fields.  In the other extreme regime which is
relevant for weaker intensities but ultra high photon energies,
perturbative multiphoton pair creation has been approached
experimentally
\cite{burke1997positron,Hu:Muller:etal:2010:Complete_QED}.  New
directions in pair production have also been proposed
\cite{schutzhold2008dynamically,dipiazza2009barrier,otto2015lifting,otto2015dynamical}
by combining both regimes 
leading to the dynamically assisted Schwinger effect.

The intermediate regime, i.\,e., nonperturbative multiphoton pair
creation\footnote{This is very similar to the intermediate domain of
  strong-field ionization between the tunneling and multiphoton regimes,
  as studied in
  \cite{Klaiber:Hatsagortsyan:etal:2015:Tunneling_Dynamics}.}  with
$\xi \approx 1$, however, is less studied
\cite{Hu:Muller:etal:2010:Complete_QED,mocken2010nonperturbative,Krajewska:Kaminski:2011:Correlations_in,Augustin:Muller:2014:Nonperturbative_Bethe-Heitler,Akal:Villalba-Chavez:etal:2014:Electron-positron_pair,li2015nonperturbative}
and a comprehensible physical picture is missing.  Despite this, many
interesting phenomena may be expected like the coherent production of
multiple pairs
\cite{schwinger1951gauge,kluger1992fermion,schmidt1998quantumkinetic},
quantum statistical influences
\cite{schmidt1998quantumkinetic,hebenstreit2009quantum}, Pauli exclusion
effects \cite{schmidt1998quantumkinetic,schmidt1999nonmarkovian} and a
mixture of signatures known from pure tunneling and multiphoton
processes
\cite{brezin1970pair,kohlfurst2014effective,otto2015dynamical,Akal:Villalba-Chavez:etal:2014:Electron-positron_pair}.
This regime requires a quantum field theoretical approach
\cite{Hebenstreit:Berges:etal:2013:Simulating_fermion} which accounts
for possible multi-pair states as well as for the temporal and spatial
variations of the electromagnetic environment.  By applying the in-out
formalism \cite{fradkin1981furry,fradkin1991quantum} in combination with
numerical solutions of the time-dependent Dirac equation
\cite{Braun:Su:etal:1999:Numerical_approach,Bauke:Keitel:2011:Accelerating_Fourier,Fillion-Gourdeau:Lorin:etal:2012:Numerical_solution},
we will analyze the fermionic field state and its coherent quantum
dynamics in the nonperturbative multiphoton regime. In comparison to
other approaches, where in most cases only the dynamics of specific
observables like the number of produced pairs is calculated, the
evaluation of the fermionic field state allows to explore all amplitudes
of different single and multi-pair states. These amplitudes become
nontrivial for space- and time-dependent external fields, because in
this case the fermionic field state is not just a product state, as is
the case for external fields which depend only on time.  Furthermore,
all possible observables of interest can be calculated once the
fermionic field state is known.  We anticipate as well the experimental
relevance as nonperturbative multiphoton pair creation may be observable
employing laser sources of lower intensities
\cite{Hu:Muller:etal:2010:Complete_QED}.

\section{Theoretical foundations}
\label{sec:theory}

\begin{figure}
  \centering
  \includegraphics[]{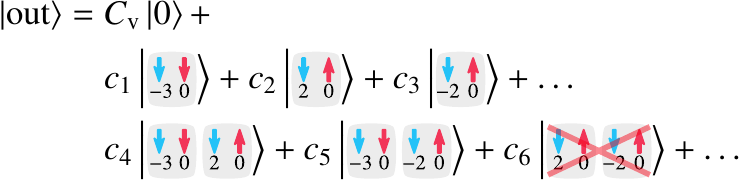}
  \caption{The vacuum evolves under the effect of a strong
    electromagnetic field into a quantum field state $\ket{\text{out}}$,
    which is a quantum mechanical superposition of the vacuum state
    $\ket{0}$ (1st row), single-electron--positron-pair states (2nd row),
    states with two electron--positron pairs (3rd row), and further
    multi-pair states.  Multi-pair states with two electrons or two
    positrons having the same quantum numbers are excluded by the Pauli
    principle (last state in 3rd row).}
  \label{fig:multi_particle_state}
\end{figure}

The pair creation probability is the central quantity in the study of
pair creation.  It is, however, a rather coarse-grained quantity.  It
does not distinguish whether only a single electron--positron pair has
been created or if the electron--positron pair is part of a
multi-particle state, which contains a larger number of
electron--positron pairs.  This fine-grained information is encoded in
the quantum field state, into which an initial vacuum state evolves
under the effect of a strong electromagnetic field.  This state is a
quantum mechanical superposition of the vacuum state and all permissible
multi-electron--positron-pair states; see
Fig.~\ref{fig:multi_particle_state}.  In this pictorial representation,
each pair's quantum numbers are represented by arrows indicating the
spin and numbers specifying the momentum.  The left arrow and number
(blue color) belong to the electron and the right (red color) to the
positron.  Due to the electron's fermionic character, the quantum
field state must be compatible with the Pauli exclusion principle, which
forbids multi-pair states, where two electrons or two positrons have the
same quantum numbers.

In strong-field quantum electrodynamics the potential of the
electromagnetic field is split into a quantized radiation field and a
classical background field.  This leads to the Furry picture formulation
of quantum electrodynamics
\cite{fradkin1981furry,fradkin1991quantum,Narozhny:Fedotov:2015:Extreme_light},
where the classical background field is treated exactly but the effect
of the quantized radiation field calculated by means of perturbation
theory.  At the onset of pair creation, where the number of created
pairs is small, it is justified to assume that the interaction between
the quantized radiation field and the quantized matter field can be
neglected completely.  In this case, the evolution of a quantum field
state in an external electromagnetic field is determined by the
Hamiltonian
\begin{equation}
  \label{eqn:qed_ham}
  \op{H}(t) = 
  \int\herm{\op{\psi}(\vec r)}\HD(\vec r, t)\op{\psi}(\vec r)\,\D^3r \,,
\end{equation}
where $\op{\psi}(\vec r)$ denotes the time-independent spinor field
operator and\footnote{Units, where the speed of light and the reduced
  Planck constant are set to unity, are employed.}
\begin{equation}
  \label{eqn:dirac_ham}
  \HD(\vec r, t) = \vec\alpha\cdot\big(-\I\vec\nabla-q\vec A(\vec r, t)\big) + m_0\beta 
\end{equation}
is the Hamiltonian density 
for a particle with charge $q$ and rest mass $m_0$.  Furthermore,
$\vec\alpha=\trans{(\alpha_x,\alpha_y,\alpha_z)}$ and $\beta$ represent
the Dirac matrices \cite{gross2004relativistic,thaller2005advanced}.
Describing the electromagnetic field via an external vector potential
$\vec A(\vec r, t)$ is justified as long as back reaction of the created
particles on the electromagnetic environment can be neglected.  The
spinor field operator $\op{\psi}(\vec r)$ can be decomposed into two
possible sets of orthonormal functions $\phiin_n(\vec r)$ and
$\phiout_n(\vec r)$ fulfilling the boundary conditions
\begin{subequations}
  \begin{align}
    \HD(\vec r, \tin) \phiin_n(\vec r) & =
    \Ein_n \, \phiin_n(\vec r) \,,\\
    \HD(\vec r, \tout) {}^{\pm} \varphi_n(\vec r) & = \Eout_n \,
    \phiout_n(\vec r) \,,
  \end{align}
\end{subequations}
where $\Ein[+]_n>0$, $\Ein[-]_n<0$ and $\Eout[+]_n>0$, $\Eout[-]_n<0$
denote the corresponding positive and negative eigenenergies, with $n$
labeling the quantum state.  The times $\tin$ and $\tout$ specify the
begin and the end of the interaction with an external electromagnetic
field, which vanishes outside the interval $[\tin, \tout]$.  The modes
$\phiin_n(\vec r)$ and $\phiout_n(\vec r)$ are chosen as free-particle
states with definite momentum, energy, and spin orientation (in the
$z$~direction) at $\tin$ and $\tout$.  Thus, in our notation the quantum
number $n$ gathers the momentum quantum number 
and the spin quantum number.

The dynamics of the quantum field state can be obtained by propagating
all basis vectors of the Hilbert space via the Dirac equation
\cite{Krekora:Su:etal:2004:Klein_Paradox,cheng2009creation,Cheng:Su:etal:2010:Introductory_review}
while including the external electromagnetic fields with their full time
and space dependence \cite{wollert2015spin}.  Computing numerically the
time propagation of the modes $\phiin_n(\vec r,t)$ yields the matrices
$\Gud+-$ and $\Gud--$ with the elements
\begin{equation}
  \Gud\pm-_{mn} = 
  \iint \herm{\phiout_m(\vec r)} 
  G(\vec r, \tout; \vec r', \tin)
  _-\phi_n(\vec r')\,\D^3r'\,\D^3r \,.
\end{equation}
As shown in Ref.~\cite{fradkin1991quantum}, the relative probability
amplitude for creating one pair composed by an electron with quantum
number $m$ and a positron with quantum number $n$ is given by
\begin{equation}
  \wmn = - [\Gud+- \GudInv--]_{mn}
  \label{eqn:omega_mn}
\end{equation}
and the vacuum-to-vacuum probability by
\begin{equation}
  \norm{C_\mathrm{v}}^2 = \norm{\det \Gud-- \,}^2 \,.
  \label{eqn:p_v}
\end{equation}
Using these quantities, the fermionic field state (including all
possible multi-pair states) at time $\tout$ can be written as
\cite{frolov78density}
\begin{equation}
  \ket{\text{out}} = 
  C_\mathrm{v} \sum_{N\{m, n\}} 
  \frac{1}{N!} \left( \prod_{i = 1}^N \wmni \right) \, \mnket \,.
  \label{eqn:state}
\end{equation}
Here, the quantum field state with $N$ pairs
\begin{equation}
  \mnket \equiv 
  \op b^\dagger_{n_1} \ldots \op b^\dagger_{n_N} \, 
  \op a^\dagger_{m_N} \ldots \op a^\dagger_{m_1} \ket{0} 
  \label{eqn:mnket} 
\end{equation}
is defined in terms of the creation operators for the electron $\op
a^\dagger_{m_i}$ and the positron $\op b^\dagger_{n_i}$ (with quantum
numbers indicated by the index sets $\{m\}$ and $\{n\}$), and the vacuum
state $\ket{0}$.  Note that the definition of $\mnket$ incorporates the
Pauli exclusion principle, because a fermionic creation operator with a
specific set of quantum numbers acting twice on some state yields zero.
The corresponding multi-pair state's probability amplitude is
\begin{equation}
  \cmn = \mnbra \text{out} \rangle \,.
  \label{eqn:coeff_nm}
\end{equation}
Accordingly, the probability that the final state $\outket$ contains $N$
pairs becomes
\begin{equation}
  c_N =  \outbra \left(
    \frac{1}{N!} \sum_{\{m,n\}} \mnket \mnbra
  \right) \outket \,.
  \label{eqn:coeff_n}
\end{equation}

Various observables can be inferred from \eqref{eqn:coeff_nm} and
\eqref{eqn:coeff_n}.  For example, the total electron spin and the total
positron spin averaged over all multi-pair states with $N$ pairs and
similarly the average total helicity may be defined respectively, as
\begin{align}
  \label{eqn:spin_n}
  s^\pm_N & = \frac{1}{c_N}\outbra \left( 
    \frac{1}{N!} \sum_{\{m,n\}} \mnket \Bigg(
      \sum_i^N  \mathfrak{s}^\pm_i  
    \Bigg) \mnbra 
  \right) \outket \,, \\
  \label{eqn:helicity_n}
  h^\pm_N & = \frac{1}{c_N} \outbra \left( 
    \frac{1}{N!} \sum_{\{m,n\}} \mnket \Bigg(
    \sum_i^N \mathfrak{h}^\pm_i  
  \Bigg) \mnbra \right) \outket \,.
\end{align}
The index ``$\pm$'' distinguishes between electrons (``$+$'') and
positrons (``$-$'') again, and $\mathfrak{s}^\pm_i$ and
$\mathfrak{h}^\pm_i$ designate the expectation values of the
\mbox{$z$-component} of the spin operator $\op{\vec\sigma}_z$ and the
helicity operator
\begin{math}
  \op h = 
  {\op{\vec{\sigma}} \cdot\op{\vec{p}}}/{\norm{\op{\vec{p}}}} 
\end{math}
of the indexed single-pair state, with $\op{\vec p}$ denoting the
momentum operator.  For example, $\mathfrak{s}^-_i$ corresponds to the
spin of the $i$th positron with the quantum number $m_i$.

\section{Elliptically polarized light beams}
\label{sec:beams}

The theoretical formalism of pair creation as outlined above is not
specific for a particular external-field configuration.  In the
following, we consider two counterpropagating monochromatic laser fields
with wavelength $\lambda$ and wave vectors equal to $\vec{k}_\pm = \pm k
\vec{e}_z$ where $k = 2\pi/\lambda$.  The electromagnetic field is
parametrized in terms of the Jones vectors \cite{jones1941new} $\ket{l}
= (\vec{e}_x + \I \vec{e}_y)/\sqrt2$ and $\ket{r} = (\vec{e}_x - \I
\vec{e}_y)/\sqrt2$, which correspond to circular left and circular right
polarization, respectively.  The electric field component of each laser
is given by
\begin{equation}
  \label{eqn:E_0_pm_complex}
  \vec{E}_{\pm}(\vec r, t) = \Re{\left(
      E \left( 
        \cos \alpha_\pm \ket{l} + \sin \alpha_\pm 
        \ket{r} \right) \e^{\I(\pm k z - \omega t)} 
    \right)}\,.
\end{equation}
The magnetic field component follows via
$\vec{B}_\pm = \vec{k}_\pm \times \vec{E}_\pm/k$, and the mean intensity
equals $\varepsilon_0 E^2/2$.  Both lasers' polarizations are determined
by $\alpha_\pm$.  Here, $\alpha_\pm$ ranges from $0$ to $\pi/2$, where
$0$ corresponds to circular left polarization (photonic spin ``down''),
$\pi/4$ to linear polarization, and $\pi/2$ to circular right
polarization (photonic spin ``up'').  For simplicity, we assume that
both plane waves have the same degree of ellipticity, which requires
$\alpha_+ = \alpha_-$ or $\alpha_+ = \pi/2 - \alpha_-$.  The helicities
of both beams are opposite to each other in the former case, whereas
they are equal in the latter.  The turn-on and turn-off of the external
field is realized by modulating the vector potential by a window
function with $\sin^2$-shaped turn-on/off of length $\Delta T$ and a
flat plateau of length $T$ in between.  Observables are always
determined after the electromagnetic field has been turned off, which is
necessary for a clear physical interpretation
\cite{blaschke2006pair,mocken2010nonperturbative,wollert2015spin}.

When the quantum field state \eqref{eqn:state} shall be determined by
numerical means, it is necessary to truncate the quantities $\Gud+-$ and
$\Gud--$ to finite matrices.  Due to the monochromatic nature of the
counterpropagating waves only quantum field modes separated by a
multiple of $\vec{k}_\pm$ couple to each other.  Thus, the Hilbert space
separates into independent subspaces \cite{wollert2015spin}.  Each of
the momentum subspaces consists of the modes with momentum $n\vec{k}_\pm
+ \vec {k}_0$, where $n$ runs over all integers and $\vec {k}_0$
identifies the subspace's momentum origin.  All parts of the fermionic
field state in Eq.~\eqref{eqn:state} corresponding to a specific
subspace can be calculated independently from others.  Furthermore,
every single subspace can be truncated to a finite number of modes due
to the fact that the coupling between different modes becomes very weak
with increasing momentum.  For simplicity, we will concentrate in the
following on the subspace with $\vec {k}_0=0$, which maximizes the pair
production probability.

\section{Multi-pair states dynamics and numerical results}
\label{sec:results}

\begin{figure}
  \centering
  \includegraphics[width=\linewidth]{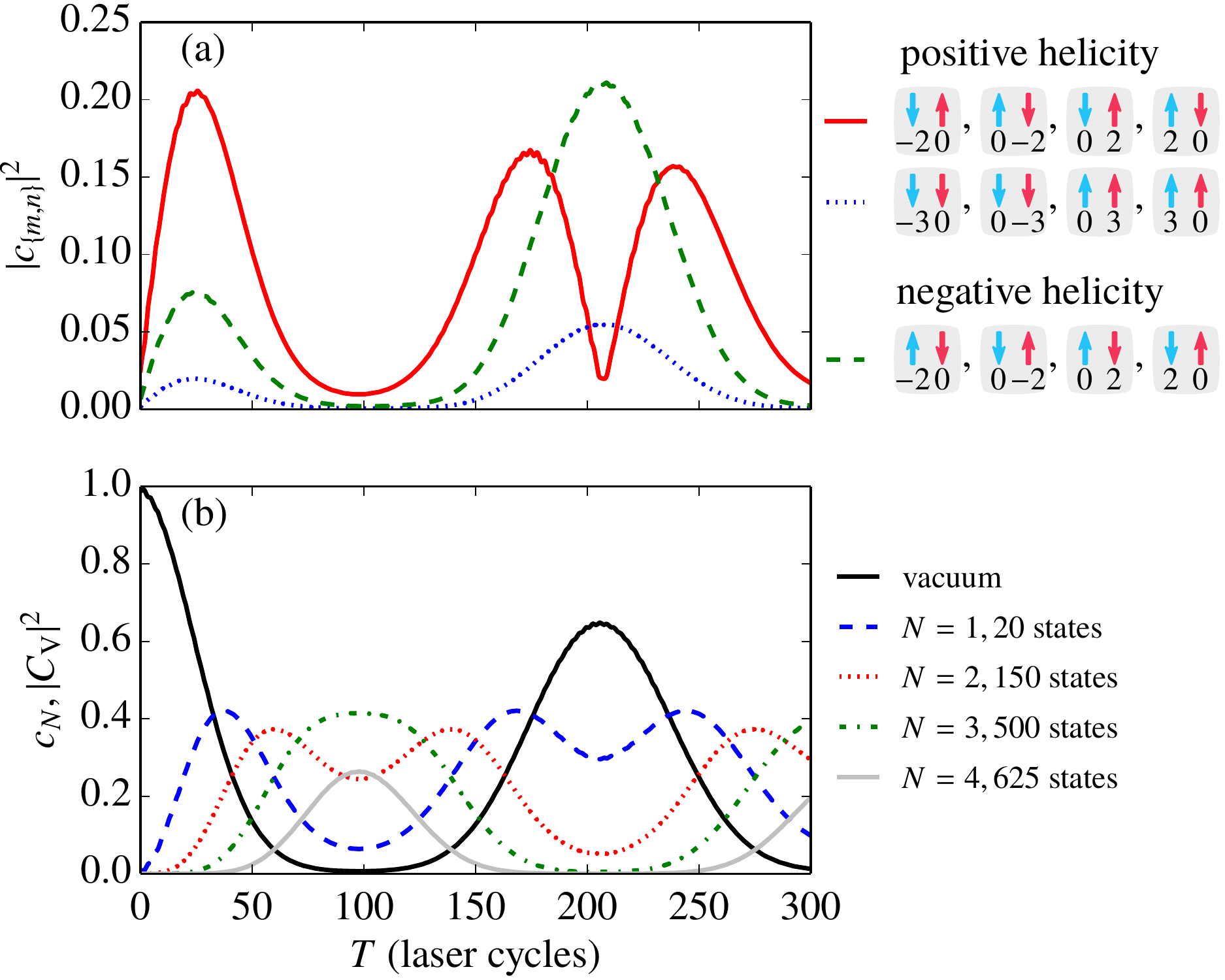} 
  \caption{(a) Evolution (corresponding to different total interaction
    periods) of single-pair states created in two counterpropagating
    laser beams of the total interaction time.  Due to symmetry reasons,
    there are always four different single-pairs with the same
    probability $|\cmn|^2$.  (b) Evolution of the initial vacuum state
    and its decomposition into different multi-pair states.  Initially,
    the states with lower pair count contribute most, but are outweighed
    later by states with higher pair count due to the high number of
    possible multi-pair states.  Laser field parameters are
    $\omega=0.746m_0$, $E=4.9\times 10^{17}\,\mathrm{V/m}$, and
    $\alpha_+=0.2\pi/4$ and both beams have the same nonzero helicity.
    Note that the electromagnetic field has been turned off before the
    probabilities $|c_{\{m,n\}}|^2$, $c_N$, and $|C_\mathrm{V}|^2$ are
    determined.}
  \label{fig:coeff}
\end{figure}

Equation \eqref{eqn:coeff_nm} provides the probability amplitudes for
states with a fixed number $N$ of created pairs and specific sets of
quantum numbers.  The resulting probabilities of the fundamental
single-pair states are presented in Fig.~\ref{fig:coeff}(a).  Our
numerical results show that only 20 single-pair states with
nonnegligible weight occur for the chosen laser field parameters and 12
of them are exemplarily displayed.  Note that all electrons/positrons
with nonzero momentum have the same helicity within each group.  The
oscillatory behavior of the probability $|\cmn|^2$ shown in
Fig.~\ref{fig:coeff}(a), which resembles Rabi flopping in atomic
physics, indicates that not only transitions from negative-energy states
to positive-energy states occurs but also the inverse process is of
importance. 

The final quantum field state also contains a certain number of
multi-pair states with nonnegligible weight, which all consist of the 20
fundamental single-pair states.  The probability to find a multi-pair
state with $N$ pairs $c_N$ and the vacuum probability $|C_\mathrm{V}|^2$
are shown in Fig.~\ref{fig:coeff}(b).  For zero interaction time, the
resulting state is in the vacuum state with probability one.  Extending
the interaction time results in a non-trivial dependence for the
probability of the final state to be in a single- or multi-pair state.
For short interaction times, it is most probable to find a single-pair
state and probabilities for multi-pair states decrease with the number
of included pairs.  This is no longer true for longer interaction times.
After an interaction time of about $100$ laser cycles, for example,
multi-pair states with $3$ or $4$ pairs are more probable than states
with $1$ or $2$ pairs.

This can be explained by combinatorics and the specific values of the
relative probability amplitudes $\wmnin$.  The probability $c_N$ is the
sum over $|\cmn|^2$ of all $N$-pair states.  The coefficient $|\cmn|^2$
for an $N$-pair state is always smaller than a coefficient $|\cpmn|^2$
for some $N'$-pair state, if the $N$-pair state includes all pairs of
the $N'$-pair state (i.\,e., $N' < N$) and all $|\wmnin|^2 < 1$.  The
coefficient $c_N$, however, can be larger than $c_{N'}$ because the
number of possible combinations of single-pair states into multi-pair
states grows exponentially with the number of pairs.  It is only limited
by the Pauli principle, which forbids to combine single-pair states with
equal quantum numbers for either the electrons or the positrons into a
multi-pair state.  Furthermore, if $|\wmnin|^2 > 1$ for two single-pair
states, then their combination has an amplitude that is larger than each
of the single pairs' amplitudes.  Nevertheless, the resulting
probability does not exceed unity because the amplitude is normalized by
$C_\mathrm{v}$; see Eq.~(\ref{eqn:state}). 

Note that in case of external electric fields, which only depend on
time, the variety of possible single-pair states is largely reduced.  As
a consequence of the homogeneity of the external electric field and the
resulting translational invariance of the Hamiltonian, only single-pair
states where the electron and the positron have the same kinetic energy
and opposite momentum are created.  This also significantly restricts
what kind of multi-pair states can be created.  For space-dependent
fields, however, single-pair states with different kinetic energies for
the electron and the positron are possible and, hence, allowed
compositions of multi-pair states are mainly restricted by the Pauli
exclusion principle when combining single-pair states into multi-pair
states.

\begin{figure}
  \centering
  \includegraphics[width=\linewidth]{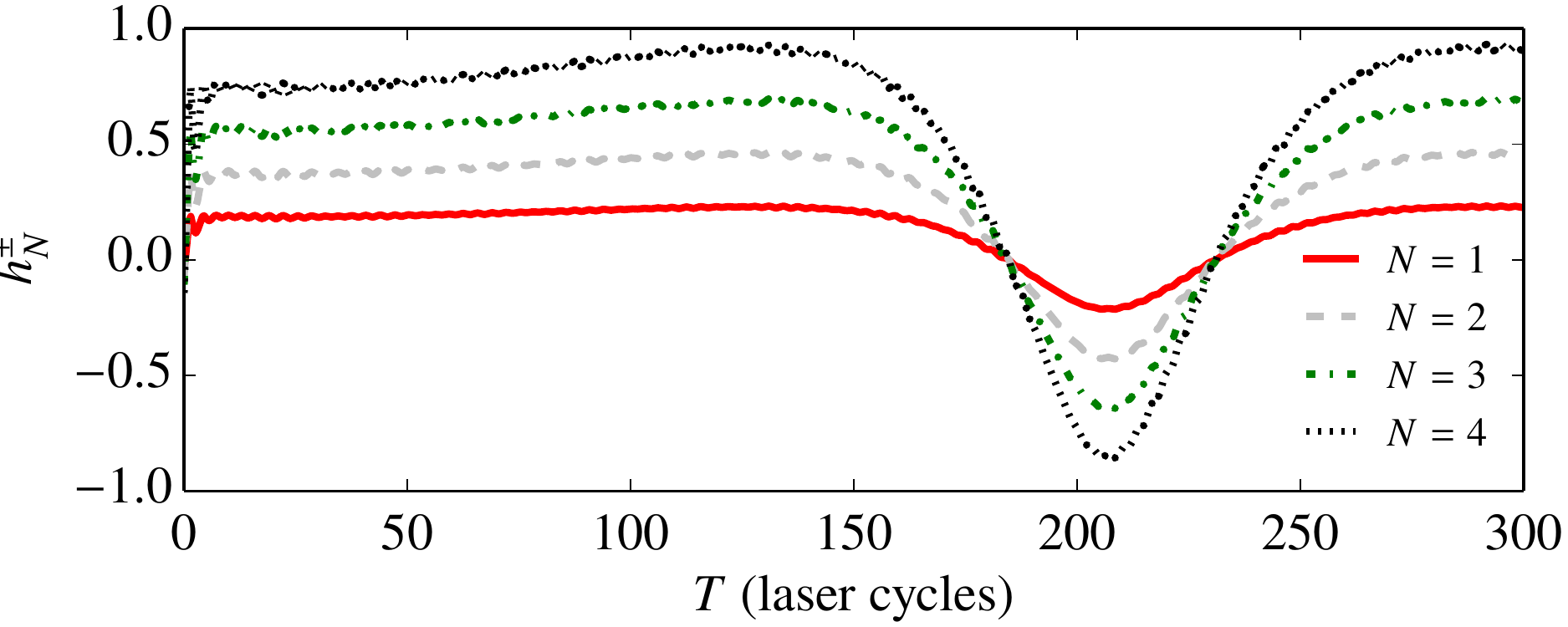} 
  \caption{Time evolution of the average total
    helicity~\eqref{eqn:helicity_n} of the produced electrons and
    positrons in different multi-pair sectors.  Laser parameters are as
    in Fig.~\ref{fig:coeff}.}
  \label{fig:helicity}
\end{figure}

The amplitudes $\cmn$ can be utilized to determine various observables
of the multi-pair states, e.\,g., the average total
spin~\eqref{eqn:spin_n} and helicity~\eqref{eqn:helicity_n}.  Depending
on whether the counterpropagating electromagnetic fields have the same
or opposite helicities two different symmetries can be found.  For
setups with the same nonzero helicities ($\alpha_+ = \pi/2 - \alpha_-$,
$\alpha_\pm\ne\pi/4$), there is a nonzero averaged helicity $h^\pm_N$
for the multi-pair states, as shown in Fig.~\ref{fig:helicity}.  The
dynamics of $h^\pm_N$ possess a sudden change from positive to negative
around an interaction time of $200$ laser cycles.  This can be
attributed to the evolution of the single-pair coefficients plotted in
Fig.~\ref{fig:coeff}(a).  Around $200$ laser cycles, states with
negative helicity (green dashed line in Fig.~\ref{fig:coeff}(a)) obtain
a larger weight compared to states with positive helicity.  For the used
setup with $\alpha_+=0.2\pi/4$, both counterpropagating laser fields'
helicities are positive but not maximal.  Each of these plane waves can
be interpreted as a superposition of an intense circular laser field
with positive helicity and a less intense circular laser field with
negative helicity.  Due to the presence of this negative-helicity field,
also multi-pair states with negative helicity are produced and their
quantum probabilities are higher than those of states with positive
helicity at around $200$ laser cycles.  Note that the definition of the
averaged helicity~\eqref{eqn:helicity_n} considers electrons and
positrons separately.  Due to symmetry reasons the average helicities
for electrons and positrons yield the same values.  Furthermore, the
average total spin~\eqref{eqn:spin_n} is zero because the
electromagnetic field has no preferred spin direction in this case.

\begin{figure}
  \centering
  \includegraphics[width=\linewidth]{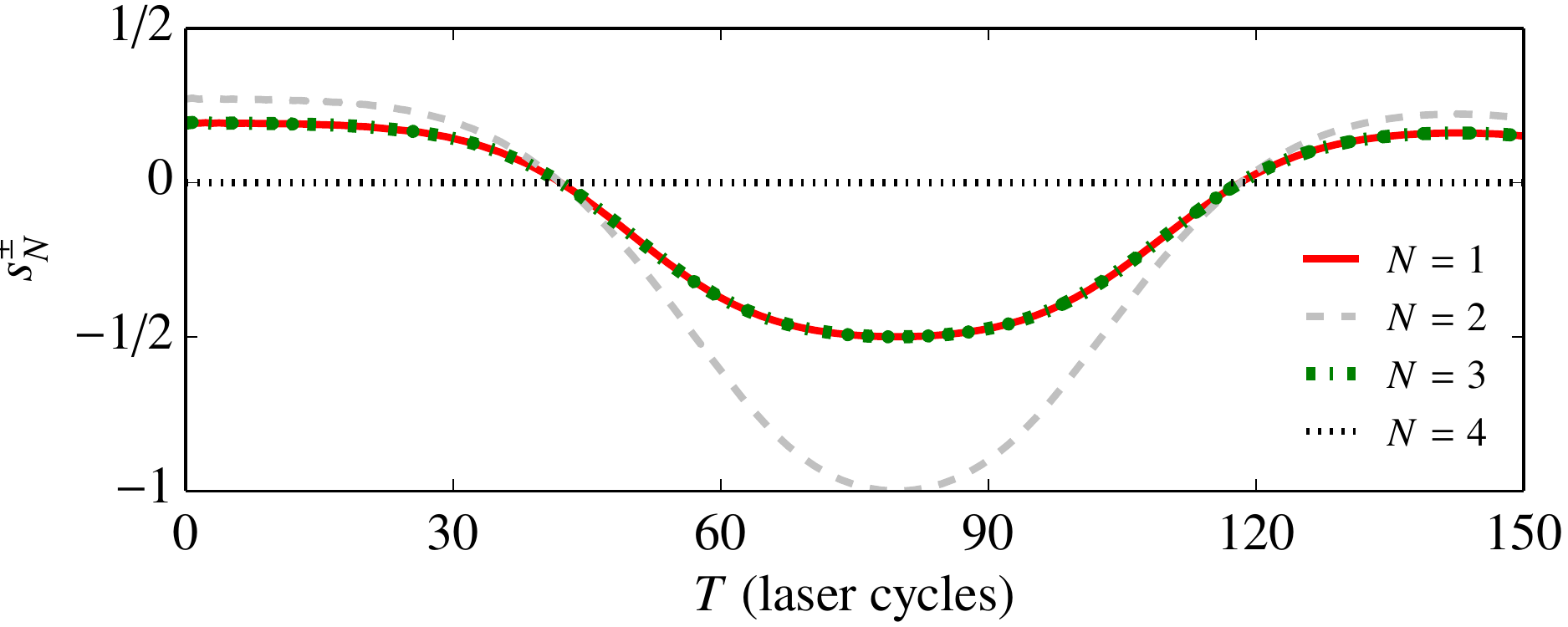} 
  \caption{Time evolution of the averaged total spin~\eqref{eqn:spin_n}
    for the produced electrons and positrons in different multi-pair
    sectors.  Nonzero spin occurs only, if both laser fields
    have opposite helicity and are not exactly linearly polarized.
    Laser field parameters are $\omega=0.4715m_0$,
    $E=3.1\times 10^{17}\,\mathrm{V/m}$, and
    $\alpha_+=\alpha_-=0.7\pi/4$.}
  \label{fig:spin}
\end{figure}

Analogously, if both laser beams have opposite nonzero helicity%
, the averaged total helicity of the created multi-pair states
vanishes.  Due to the now preferred spin direction of the laser field,
the averaged total spin $s^\pm_N$, defined in Eq.~(\ref{eqn:spin_n}),
can take nonzero values which in turn may lead to spin-polarized
electron--positron pairs \cite{strobel2015semi,wollert2015spin}.  The
averaged spin changes its sign depending on the interaction duration for
the same reasons given above for the helicity, see Fig.~\ref{fig:spin}.
The averaged spin values for the $N=3$ multi-pair sector are the same as
for the single-pair states with $N=1$ and, furthermore, the averaged
spin values for the $N=4$ sector is exactly zero.  Both peculiarities
can be explained by the constraints due to the Pauli principle when
combining single-pair states into multi-pair states.  For example, all
permissible \mbox{4-pair} states consist of exactly two pairs with the
electron and the positron having spin up, and two pairs with the
electron and the positron having spin down, which yields a zero averaged
spin value.  A hypothetical state of $N=4$ pairs with a nonzero total
spin, where $3$ or $4$ of its single pairs have the same spin for
electron (positron), is not allowed because at least one of the single
pairs would share the same quantum numbers with another pair.

\section{Conclusions} 
\label{sec:conclude}

A quantum field theoretical treatment of the matter field has been
put forward, which yields by numerical means the fermionic quantum
field state that results after the action of a strong electromagnetic
field.  It contains all information about the spectrum of created pairs,
including single pairs but also multi-pair configurations.  This quantum
field state can be determined by solving numerically the time-dependent
Dirac equation for all basis vectors of the corresponding Hilbert space.
This approach opens the door to study the complex dynamics of pair
creation including the creation of multi-electron--positron states and to
characterize also correlations between various multi-pair states.  It is
particularly useful for studying nonperturbative multiphoton pair
creation at $\xi\approx 1$.  It is, however, neither restricted to
$\xi\approx 1$ nor to $\omega\approx m_0$, which was chosen here to keep
the computational demand small.  Furthermore, the combination of
high-intensity lasers with accelerated electrons may produce effective
field strengths close to the critical field
\cite{burke1997positron,bamber1999studies,dunne2009newstrong}, which is
a scenario that can also be treated by the approach as proposed here
with minor modifications.

A striking feature of this method is that it yields the probabilities of
detecting specific multi-pair states and it shows how these
probabilities evolve during the course of interaction with the
electromagnetic environment.  Furthermore, the polarization and symmetry
of the classical laser field persist in the quantum realm for the
produced particles.  Finally, also pure quantum effects like the Pauli
exclusion become apparent.

\section*{Acknowledgements}

We acknowledge helpful discussions with Antonino Di Piazza and Sebastian
Meuren.

\bibliography{refs}

\end{document}